# Deep learning-based framework for cardiac function assessment in embryonic zebrafish from heart beating videos

Amir Mohammad Naderi[a], Haisong Bu[b], Jingcheng Su[a], Mao-Hsiang Huang[c], Khuong Vo[d], Ramses Seferino Trigo Torres[e], J.-C. Chiao[f], Juhyun Lee[g], Michael P.H. Lau[h], Xiaolei Xu[b], and Hung Cao[a,e,h]

[a] Department of Electrical Engineering and Computer Science, University of California, Irvine, CA

[b] Department of Biochemistry and Molecular Biology/Department of Cardiovascular Medicine, Mayo Clinic Rochester, MN

[c] Department of Biomechatronics Engineering, National Taiwan University, Taipei, Taiwan

[d] Department of Computer Science, University of California, Irvine, CA

[e] Department of Biomedical Engineering, University of California, Irvine, CA

[f] Department of Electrical and Computer Engineering, Southern Methodist University, Dallas, TX

[g] Department of Bioengineering, University of Texas, Arlington, TX

[h] Sensoriis, Inc., Edmonds, WA

Correspondence

Hung Cao, Ph.D.

Assistant Professor of Electrical and Biomedical Engineering

University of California, Irvine

Irvine, CA 92697

E-mail: hungcao@uci.edu

**Abstract**

Zebrafish is a powerful and widely-used model system for a host of biological investigations including cardiovascular studies and genetic screening. Zebrafish are readily assessable during developmental stages; however, the current methods for quantification and monitoring of cardiac functions mostly involve tedious manual work and inconsistent estimations. In this paper, we developed and validated a Zebrafish Automatic Cardiovascular Assessment Framework (ZACAF) based on a U-net deep learning model for automated assessment of cardiovascular indices, such as ejection fraction (EF) and fractional shortening (FS) from microscopic videos of wildtype and cardiomyopathy mutant zebrafish embryos. Our approach yielded favorable performance with accuracy above 90% compared with manual processing. We used only black and white regular microscopic recordings with frame rates of 5-20 frames per second (fps); thus, the framework could be widely applicable with any laboratory resources and infrastructure. Most importantly, the automatic feature holds promise to enable efficient, consistent and reliable processing and analysis capacity for large amounts of videos, which can be generated by diverse collaborating teams.

**Keywords**: Zebrafish, Heart Disease, Cardiomyopathy, Deep Learning, Ejection Fraction

## 1. Introduction

Despite extensive research and medical expenditures, cardiovascular disease continues to be the leading cause of mortality and morbidity in the modern world [1]. Among animal models used in cardiovascular research, zebrafish (*Dario rerio*) has been proven to be the premier model for studies of developmental genetics and functional genomics owing to their conserved genome, small size, low-cost for maintenance, short generation time, and optical transparency, just to name a few [2]. In addition, zebrafish cardiac physiology shows similar phenotypes to that of humans [3] and the time-lapse videos of the heart development can be easily acquired [4]. All these make zebrafish an ideal choice to investigate cardiac development, congenital heart disease as well as therapeutic potentials.

Dilated cardiomyopathy (DCM) is a hereditary, progressive disease, which eventually leads to heart failure [6]. Thus, it is very important to evaluate the early cardiac functions associated with DCM. Dozens of pathogenic genes have been found in the genetic studies of cardiomyopathy, and the incidence rate of DCM is about 1/250 [7]. Titin truncated variants (TTNtv) are the most common genetic factor in DCM, accounting for 25% of DCM cases [8]. Therefore, we have recently restated the allelic heterogeneity in zebrafish segments and established a stable mutation system to systematically and accurately assess the cardiac functions of mutant zebrafish. In order to study the mechanobiology of induced defects of these disease models, heart functions need to be reliably assessed [9]. This effort needs a thorough knowledge of blood flow patterns as well as hemodynamics. Furthermore, in studies on cardiogenesis for screening distinct roles of different genes in mediating heart development and cardiac functions, investigating cardiomyocyte sizes and numbers resulting in measures of the ventricular chamber volume was usually utilized [10]. To this end, a systematic and simple approach for quantifying cardiac functions in zebrafish embryos would provide important insights into of the development of phenotypes and disease. The common quantitative indices are Ejection Fraction (EF) and Fraction Shortening (FS) which are different measures of the heart's muscular contractility.

The embryonic zebrafish (up to ~3 days post fertilization - dpf) are transparent with decent visibility of internal organs, including heart and blood circulation. Thus, bright field microscopic videos can be used for quantification of heart mechanism and morphology at this stage [11]. Usually, two dimensional (2D) videos for cardiovascular analysis are recorded. Then, continuous changes in ventricular wall position throughout the cardiac cycle would be tracked by first identifying a linear region of interest for the borders of the ventricle [11]. Structural analysis of the zebrafish heart is based on taking 2D images at specific time points to measure chamber dimensions. However, in conventional approaches, researchers have to manually label the ventricle, find the End Systolic (ES) and End Diastolic (ED) frames, and then derive the desired parameters, such as EF and heartrate (HR). Till date, most of the reported work only dealt with simple detection of heartrate, such as via edge tracing [12]. Nasrat *et al.* presented a method for semi-automatic quantification of FS in video recordings of zebrafish embryo hearts [13]. Their software provides automated visual information about the ES and ED stages of the heart by displaying corresponding-colored lines into a motion-mode display. However, the ventricle diameters in frames of ES and ED stages are marked manually, and then the FS is calculated. This will be extremely tedious, time consuming and inconsistent when segmentation is done manually for a large number of frames. Akerberg *et al.* proposed a Convolutional Neural Network (CNN) framework that automatically segments the chambers from the videos and calculates the EF [14]. Nevertheless, particular transgenic animals expressing the myocardial-specific fluorescent reporter and hi-end fluorescence microscopes were used, which cannot be widely applicable for the research community, especially those without access to transgenic lines or fluorescence microscopes. Additionally, Huang *et al.* showed that transgenic expression of fluorescence protein can cause dilated cardiomyopathy [15], as high levels of expression of some foreign proteins affect the myocardium. Further, more importantly in the work reported by Akerberg and colleagues, frames from only 4 videos have been used which can result in overfitting in cases where the features of the video like the position of the fish, lighting, or the focus of the lens on the ventricle are different comparing to the training set. Zebrafish in the videos can have different sizes and the focus on the heart can be different in each video. In manual segmentation, the ventricle could occasionally be partially masked. Therefore, in order to have a framework with the ability to estimate these masked spots accurately, the dataset should include a variety of videos with different settings. However, their work has raised the optimism of applying machine learning to this problem.

U-net, a symmetric convolutional neural network architecture, could be an ideal option since it is specifically created for biomedical image segmentation [16]. A similar architecture has been employed by Decourt *et al.* to segment the human left ventricle from magnetic resonance imaging (MRI) images [17]. The main idea of the U-net is to complement a traditional contracting network by successive layers, where pooling operations are replaced by up-sampling operators. Besides, a successive convolutional layer can then be trained to assemble a precise output based on this information [16]. The training of the network uses the original image as an input and the mask of the corresponding image as the output and the objective is to minimize the error of the estimation and the mask.

In this work, our *Zebrafish Automatic Cardiovascular Assessment Framework* (ZACAF) was developed to analyze heart beating videos of zebrafish embryos simultaneously and quantify specific cardiac functions, namely FS and EF. A deep learning model has been trained using 50 videos of wildtype and mutant zebrafish. Additionally, several image processing techniques have been applied to the videos to investigate the effectiveness and then the best one was chosen to preprocess the data before training. We then evaluated the performance of our framework with the wildtype as well as the established TTNtv mutant fish having a condition of DCM [12]. Finally, in-depth discussion and future directions are presented.

## 2. Methods

### 2.1 Experimental animals

Zebrafish (*Danio rerio*; WIK strain) were maintained under a 14 h light/10 h dark cycle at 28.5°C. All animal study procedures were performed in accordance with the Guide for the Care and Use of Laboratory Animals published by the U.S. National Institutes of Health (NIH Publication No. 85-23, revised 1996). Animal study protocols were approved by the Mayo Clinic Institutional Animal Care and Use Committee (IACUC #A00002783-17-R20).

### 2.2 Video imaging of beating zebrafish hearts at embryonic stage

Zebrafish in the embryonic stages were anesthetized using 0.02% buffered tricaine methane sulfonate (MS222 or Tricaine) (*Ferndale,* Washington, US) for 2 minutes, and then placed lateral side up with the heart facing the lower-left corner. The specimens were held in a chamber with 3% methylcellulose (*Thermo Fisher Scientific*, Massachusetts, US). The videos were recorded using a Zeiss Axioplan 2 microscope (Carl Zeiss, Oberkochen, Germany) with a 10X lens and differential interference contrast (DIC) capacity. The used Zeiss' Axiocam 702 mono Digital Camera 426560-9010-000 records videos with 60 fps; however, using the Zeiss computer software videos get stored in 5 fps, 10 fps, and 20 fps. Video clips were processed using ImageJ for manual quantification of cardiac functional indices including heart rate and fraction shortening, as detailed in the next sections.

**2.3 Cardiac function assessment**

Fraction Shortening (FS), one of the measures of ventricular contractility, can be calculated from ventricular diameters (Short-axis) at end-diastole (ED) and end-systole (ES) (Dd and Ds, respectively) as follows [11]:

$$FS = \frac{(D_d - D_s)}{D_d} \quad (1)$$

By assuming a prolate spheroidal shape for the ventricle, the following volume formula can be used: $Volume = \frac{1}{6} \times \pi \times D_L \times D_S^2$ (2) where $D_L$ and $D_S$ are long and short-axis diameters of ventricle from 2D static images as shown in **Figure 1**.

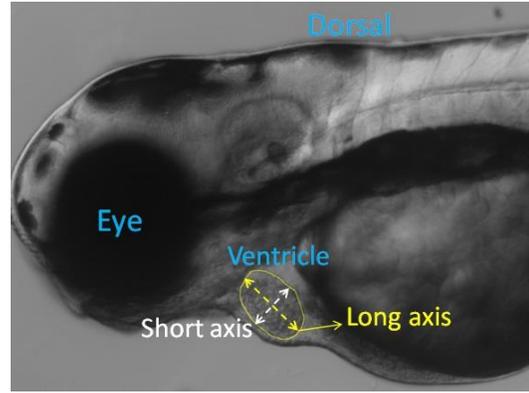

**Figure 1.** A frame in a video recorded from a 3-dpf zebrafish with segmentation for ventricle border and long and short axes.

One extremely important index in quantifying the heart mechanisms is ejection fraction (EF). It is defined as the fraction of blood ejected from the ventricle with each heartbeat and can be calculated using the following formula:

$$EF\% = \frac{(EDV - ESV)}{EDV} \times 100\% \quad (3)$$

where EDV and ESV are volumes at ED (EDV) and ES (ESV), respectively. Since measurement of the volume of the ventricle is not possible with 2-D videos, researchers usually use the area as an estimation. Finally, heartrate (HR) could be determined by measuring the time between two identical successive points (*i.e.*, ED or ES) in the recorded images [11]. The range of EF is 50%-70% for healthy zebrafish and it is one of the important indications to diagnose heart failure. EF reflects the function of the ventricular systolic pump. The stronger the myocardial contractility, the higher the stroke volume and EF will be. Therefore, in patients with heart failure, the left ventricular ejection fraction of the heart will be significantly reduced.

**2.4 Automated quantification of cardiovascular parameters using image processing**

For automatic quantification of important cardiovascular parameters like EF or FS from the microscopic videos, the ventricle of the zebrafish needs to be segmented. Several image processing methods have been employed as an effort to identify the edges of the ventricle in the videos; including edge detection, background subtraction, color filtering, and, histogram-based segmentation. In edge detection, the Canny algorithm was used which has a multi-stage algorithm to detect a wide range of edges in images [18]. In background subtraction, continuous frames from a video would be subtracted from each other to find the moving objects. Considering that most of the fish body is static and the only pixels moving in the video belong to blood cells and the heart; thus, the static pixels can be removed. Lastly, the idea behind color filtering and histogram-based segmentation is to identify the ventricle due to its distinct color or gray intensity. In manual histogram thresholding, after plotting the gray scale histogram of the image, those peaks and valleys in the histogram are used to locate the clusters in the image [19]. Otsu's algorithm performs automatic image thresholding by finding a single intensity threshold that separates pixels into two classes of foreground and background [20]. Finally, contrast limited AHE (CLAHE) is a variant of adaptive histogram equalization which over-amplifies the contrast on small regions in the image [21]. Here, the mentioned methods have been implemented not only to compare with the approach using deep learning described in the next

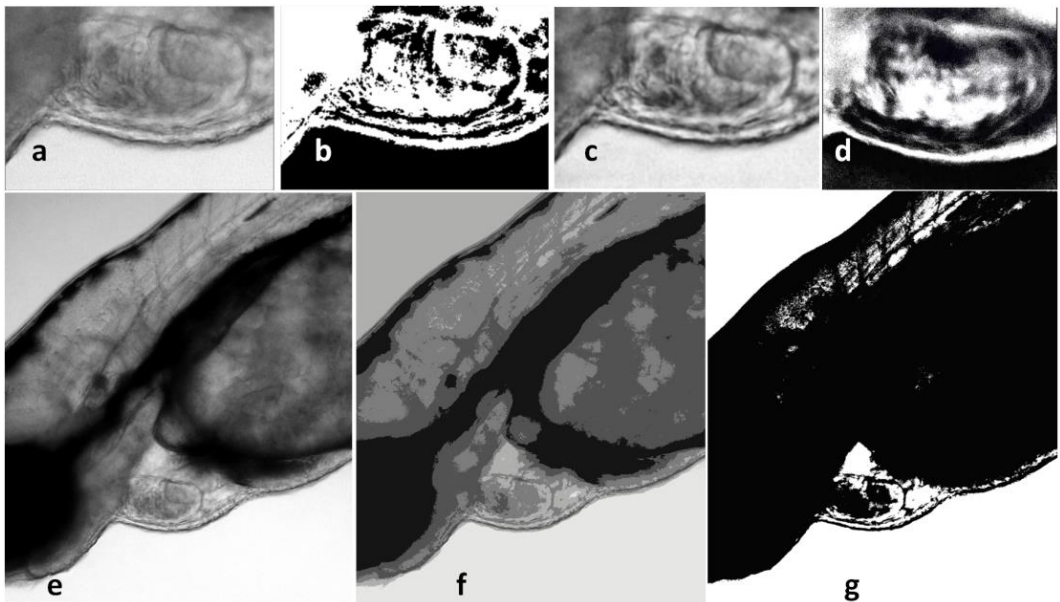

**Figure 2. Ventricle segmentation using different methods.** Panel **a-d**: A frame from the video of a 3 dpf zebrafish with 40X zoom undergoing different HBS algorithms. **a**. Original frame. **b**. Manual histogram thresholding. **c**. CLAHE. **d.** Otsu thresholding. Panel **e-g**: A frame from the video of a 3 dpf zebrafish with 10X zoom undergoing GMM and K-means approaches. **e**. Original frame **f**. GMM. **g**. K-means.

section but also to use them for preprocessing. All these are shown in **Figure 2, a-d** panels**.**

Those abovementioned methods, namely edge detection, color filtering and background subtraction are not robust with different videos, since ventricle edges might have multiple shades of gray. Therefore, we also attempted to use machine learning approaches to compare. First, unsupervised learning segmentation methods like K-means and Gaussian mixture model (GMM) were applied to the videos. As can be seen in **Figure 2e, f** and **g**, although these methods improve the visibility of the ventricle borders, the automatic segmentation of the heart is not possible. Moreover, many of the unnecessary information (pixels) in the image, particularly in the image generated using K-means, are still remaining.

## 2.6 U-net-based deep learning approach

**Figure 3** illustrates the architecture of the proposed U-net model with details. The network consists of a contracting path and an expansive path, which gives it the U-shaped architecture.

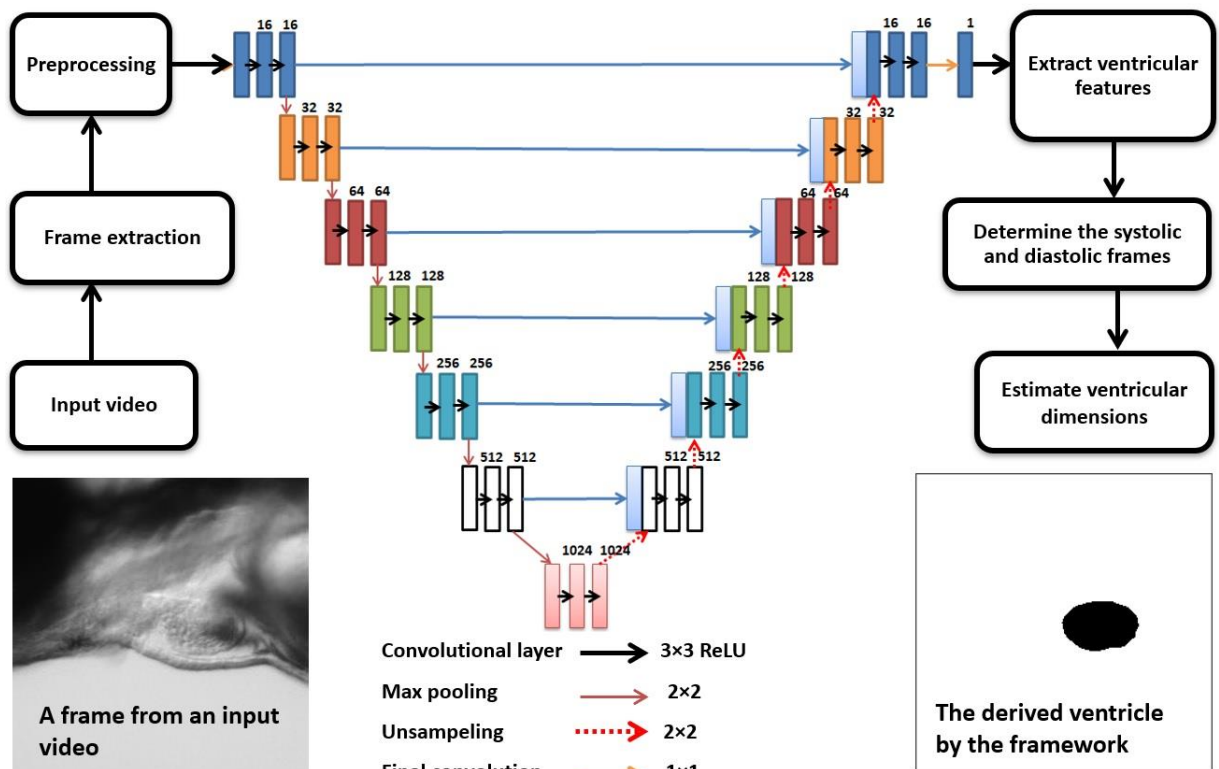

**Figure 3. The process flow and the U-net architecture.** Each rectangle represents a layer and the number above it shows the number of the neurons inside. A trained model can estimate a mask of the ventricle from all the extracted frame of the input video. When all the frames have a predicted mask, by determination of ES and ED frames, important cardiac indices like EF, FS, and stroke volume can be automatically calculated and saved in a desired format.

The contracting path is a typical convolutional network that consists of repeated convolutions, each followed by a rectified linear unit (ReLU) and a max-pooling operation. Dropouts have been used to prevent overfitting. The architecture has been optimized to obtain the best result. For training, NVidia's T4 GPU from Google Collaboratory was employed. To evaluate the model, the most commonly used loss functions for semantic image segmentation were deployed, namely Binary Cross-Entropy and Dice loss function. Cross-entropy can be defined as a measure of the difference between two probability distributions for a given random variable or set of events. It is extensively used for classification problems and since segmentation is the classification in pixel level, cross-entropy has been widely used. Binary Cross-Entropy is defined as:

$$Loss_{BCE}(y, \hat{y}) = -(y\log(\hat{y}) + (1-y)\log(1-\hat{y})) \quad (4)$$

where y is the true value and $\hat{y}$ is the predicted outcome.

The Dice coefficient is a commonly used metric in computer vision problems for calculating the similarity between two images. In 2016, it has also been adapted as a loss function, namely Dice Loss [22].

$$Loss_{Dice}(y, \hat{y}) = 1 - \frac{2y\hat{y}+1}{y+\hat{y}+1} \quad (5)$$

The U-net model has been trained with both models and the performance has been assessed using validation and test sets. Further, calculation of EF has been also evaluated using both aforementioned loss functions.

*a. Dataset*

A training dataset was created employing the raw microscopic videos of zebrafish containing 800 pixel-wise annotated images. 50 videos of the lateral view from multiple 3-dpf zebrafish were analyzed for creating the dataset. 10 of these videos are from the TTNtv mutant line. From each video, 10 to 30 frames were extracted. Each training set has a frame from the video and a mask manually created showing only the ventricle with ImageJ software. After creating the masks, all image and mask sets have been organized into folders. Each set has two folders inside, one for the original extracted frame and the other for its corresponding mask. Finally, all sets were shuffled to avoid overfitting. The validation set with the 10% of the size of the data has been split from the dataset prior to training.

*b. Preprocessing*

In the preprocessing stage, a region of interest is defined knowing all recordings have the same positioning for the zebrafish. Although this cropping improves the accuracy by removing unnecessary information, it can be avoided to make the framework robust to different video types. Additionally, a sharpening filter accomplished by performing a convolution between a custom weighed kernel and an image is used to make edges more visible. After training, the U-net architecture was able to predict the ventricle segment. The model has been trained several times by applying the mentioned image processing methods to the training images. The method with the best results was CLAHE thresholding which was added to the preprocessing section.

*c. Quantification of the diameters of the predicted ventricle*

The diameters of the ventricle are measured for all extracted frames automatically with contour tool from OpenCV (an open-source computer vision library). The maximum and minimum measured areas of the ventricle in different frames show the ES and ED stages, respectively. Using the measurement of ES and ED frames, we can calculate the ejection fraction (EF), fractional shortening (FS), and stroke volume (SV). Also, the time between two ES (or ED) frames could be used to derive heartrate (HR). The predicted ventricle is assumed to be an ellipsoid. For quantification of EF, the area of the ventricle can be used (**Eq (3)** above) by counting the pixels inside the predicted shape. Since the frames are 2D, we are estimating the ventricle volume to its area. For FS, measurements of the short axis in ES and ED frames are needed. As the ventricle is not a perfect ellipsoid, estimation of the short and long axes can be carried out in two different ways. In the first method, an ellipsoid could be fitted in the predicted shape and then the axis of the fitted ellipsoid would be measured. The second way is to find the longest line as the long axis of the estimated ellipsoid which could be found in the geometrical shape; then, the short axis of the ellipsoid is the short axis of the ventricle.

*d. Graphical User Interface (GUI)*

This framework was developed in Python and thus, for researchers who are not familiar with programming, working with it can be challenging. To address this, a Graphical User Interface (GUI) has been designed to provide a user-friendly interface to facilitate the process for researchers. Moreover, after training the U-net, the trained model can be saved which means the most computationally heavy part could be done only once. The GUI saves the output files in the

CSV format, along with information about EF, FS, diameter readings of the area, short and long axis, and frame numbers. Therefore, the data for each video can be easily accessed at any time, and anywhere with the expandable cloud feature. Our ZACAF provides an end-to-end interface to researchers to automatically calculate, classify, and record various cardiac function indices reliably. ZACAF is able to work with multiple videos at the same time and output the results in the fraction of the time compared to that of manual segmentation. The deep learning model in the ZACAF can easily be updated and optimized with a new model and data.

**2.7 Quantitative comparison of approaches**

In this framework, our objective is to predict the geometrical shape identifying the ventricle with high accuracy in terms of its position, size, and shape with the ground truth. Since the manually created masks are considered as the ground truth, we would expect the predicted shape and the manual mask to be identical or close to them. In semantic image segmentation, the most commonly-used metrics include pixel-wise accuracy, Dice coefficient, and Intersection over Union (IoU).

a. *Pixel-wise Accuracy*

In this work, since the mask indicating the ventricle is either white or black, there are only two classes so we can use the binary case of pixel accuracy. The accuracy is defined as the percent of pixels classified correctly as

$$pixel - wise\ Accuracy = \frac{pixels\ classified\ correctly}{All\ pixels} \qquad (6)$$

b. *Dice coefficient*

Dice coefficient is an extensively used indicator to elaborate the similarity of two objects. It ranges from 0 to 1 in which 1 means perfectly matched or completely overlapped. For a binary case, the coefficient is calculated as

$$Dice = \frac{2|(A \cap B)|}{|A|+|B|} \qquad (7)$$

where A is the predicted image and B is the ground truth (manually created mask).

c. *Intersection over union*

It is also known as the Jaccard Index which is simply the area of overlap between the predicted segmentation and the ground truth divided by the area of union between the predicted segmentation and the ground truth. This metric ranges from 0–1 with 0 signifying no overlap and 1 signifying perfectly overlapping segmentation. For the binary case, it can be calculated as:

$$J = \frac{|A \cap B|}{|A \cup B|} \tag{8}$$

In this work, all three of the mentioned metrics have been used to show the performance of the framework.

## 3. Results

### 3.1 Assessment of the accuracy of the framework with the defined metrics

The model's performance can be seen in **Figure 4**. The model has been trained with two loss functions discussed in section **2.6** and the best results with parameter tuning are illustrated. The aforementioned metrics resulted in 99.1% for pixel-wise accuracy, 95.04% for Dice coefficient, and lastly 91.24% and for the IoU. All mentioned metrics are evaluating the best performing model that had a Dice loss function with an Adam optimizer and a 0.001 learning rate. Validation split was 10% which means 80 sets. Following the training, we visually assessed the framework's ability to correctly segment ventricular chambers and also the periodic pulsating movement of it within series of frames of a test video. This process was used in parameter tuning for the deep learning model.

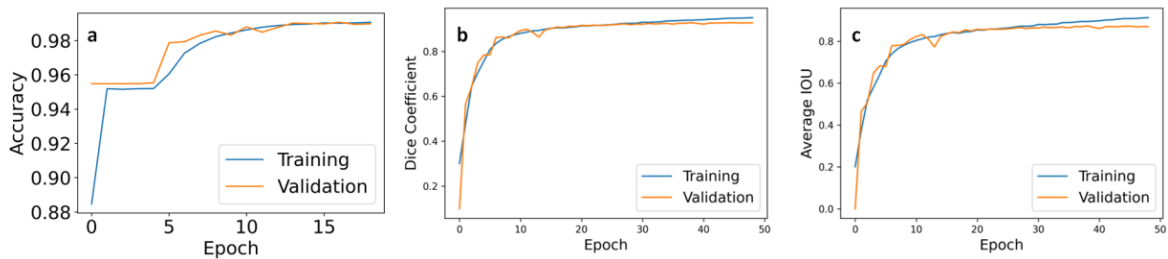

**Figure 4. The proposed model's performance plotted with the metrics commonly used in semantic image segmentation. a.** Pixel-wise accuracy **b.** Dice coefficient **c.** IoU metric. This plot shows the performance of the framework with the training and validation sets during the process of training of the deep learning model.

### 3.2 Assessment of the performance of the framework for EF

The framework was evaluated by comparing the results obtained by manual assessment of EF from an experienced biologist with those using the software since one of the primary purposes of this framework is EF calculation. In this calculation, finding the area in all frames of a video is important because we want to find the ED and ES areas. Hence, assessment should involve the series of frames in a test video rather than having random images in a validation set. For this reason, we assess the performance of ZACAF with EF calculation. First, 8 videos of wildtype zebrafish embryos and another 8 from TTNtv mutant embryos were used as the input to the framework. These videos are the test set and have not been used in the training. Second, manual processing and estimation were performed for each video to derive EF by an expert to use as the ground truth. The program saves the predicted ventricle masks for each and every frame of a video and the ED and ES frames are simply the frames with maximum and minimum area of the segmented ventricle respectively. After automatically finding ES and ED frames, the EF of the fish in the input video would be calculated and saved in a CSV file along with other indices calculated. The averages of absolute errors and standard deviations for the calculated EF of the 8 wild type test videos comparing to expert's manual calculation were 6.13% and 3.68% respectively. As ED and ES frames are the most important parameters to quantify cardiovascular indices, we plotted the correlation of the automated and manual measurements (**Figure 5**).

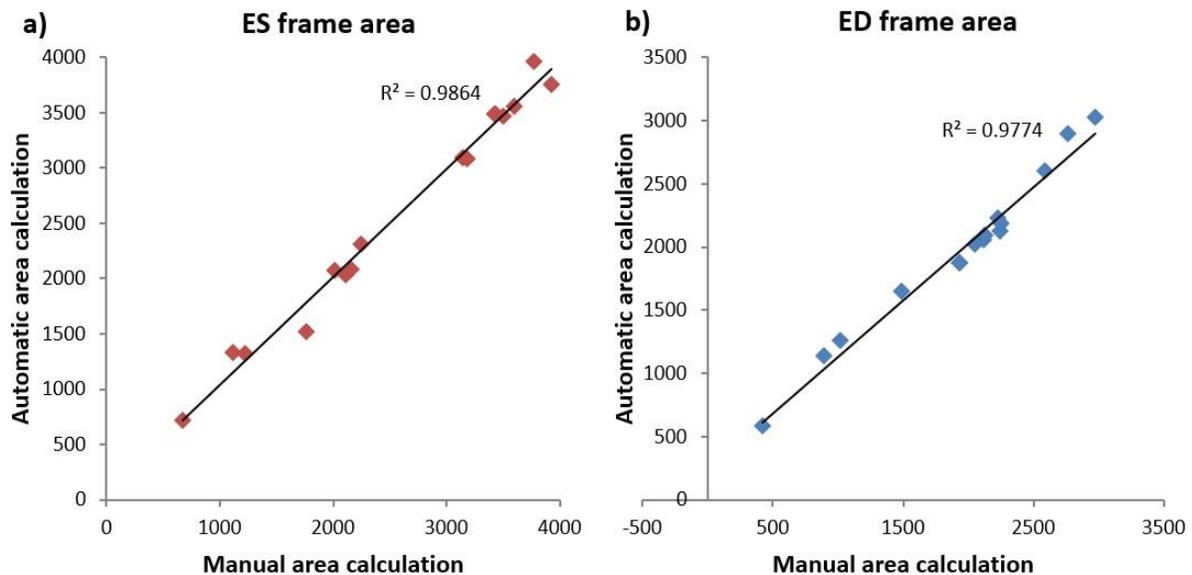

**Figure 5.** After finding and measurement of the ventricle area in ED and ES frames of 8 wild type and 8 TTNtv mutant fish with both manual and automated methods, the results are demonstrated in a correlation plot. Linear relation of the measurements with slopes close to 1 shows the accuracy of the ZACAF. (**a**) ES frame area. (**b**) ED frame area.

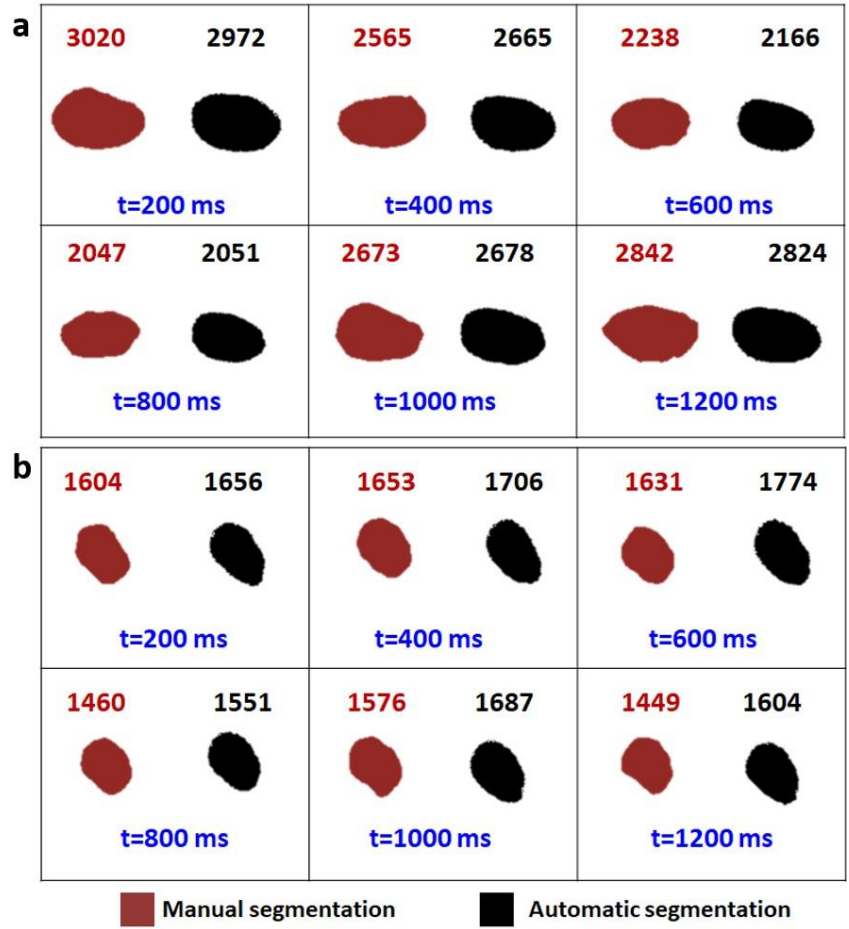

**Figure 6. Validation of U-net image segmentation framework.** The sequential frames from a wild type zebrafish recorded video with fps of 5 are extracted. The respective ventricle mask of each frame is shown in each panel via manual and automatic segmentation. The area of each ventricle is measured and written above its own box. Considering the fps of the videos and the average heart rate of the zebrafish 6 consecutive frames have been shown in this figure to ensure having at least one full cycle.

**Figure 6** presents the comparison of manual and automatic segmentation of the ventricle in 6 continuous frames to cover an entire cardiac cycle for both wild type (**a**) and TTNtv (**b**). In manual segmentation, measures were done using freehand selection tool in the ImageJ software.

**3.3 Comparison of video recording features**

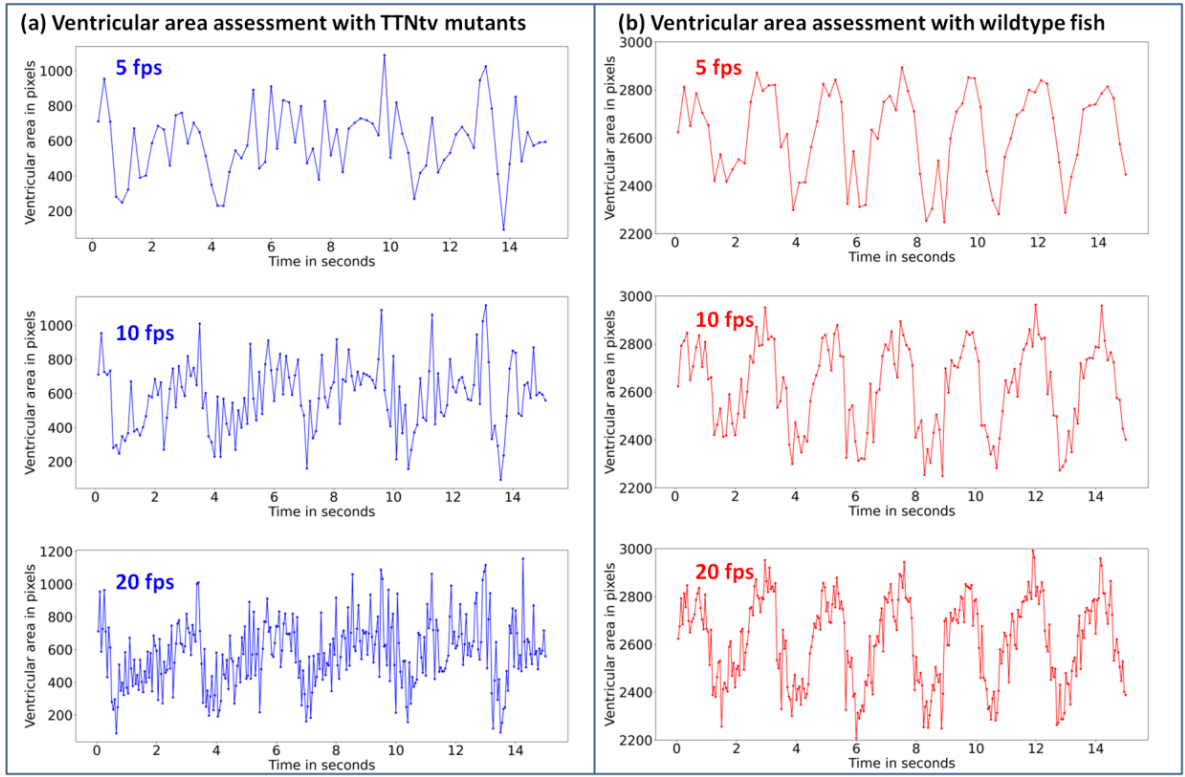

**Figure 7. Ventricular area over time.** The area is measured by counting the number of pixels inside the desired shape. Comparison of the ventricle areas from a sequence of extracted frames from 5 fps, 10 fps and 20 fps videos, respectively. Sequence of continues frames from a test video have been segmented automatically. The measured ventricle area in continues frames plotted with the same video mapped to different frame rates. **a)** In a TTNtv mutant video and **b)** in a wild-type video.

We compare the performance of the program for input videos with different frame rates, *i.e.* frames per second or fps. **Figure 7a** and **b** show the ventricular area changes over time plotted for videos with 5, 10, and 20 fps for a wild type video and a TTNtv mutant video selected from the test set. Further, from these plots, HR and stroke volume (SV) can be simply calculated. The videos with different frame rates were derived from the raw recording of the camera which records videos with 60 fps. The purpose here was to investigate the importance of the fps value in assessing the indices.

## 4. Discussion

Conventional image processing methods such as background subtraction, HBS, and edge detection have been widely used for segmentation showing excellent efficiency and efficacy; however, in this case, as can be seen in **Figure 2**, automatic segmentation of the ventricle is still not possible. Past work involving machine learning algorithms have showed promise with the proved semi-automatic feature, however requiring specific fluorescence videos. Our framework can be employed to assist researchers to quantify the cardiac functions and parameters of studied zebrafish with minimum manual engineering efforts. In EF derivation, counting the pixels is more relevant and accurate than finding the long axis which can be complicated since the ventricle is not a perfect ellipse. Further, the tool that most researchers use in the ImageJ software is a freehand ruler which could introduce inaccuracy, especially with the small size of heart chambers.

Additionally, manual segmentation is not consistent. Segmentation of the ventricle in these videos is a very difficult task even manually. The small size, ambiguous edges, and partial obstruction of the heart in the videos can also add complication to manual detection. We have investigated this quantitatively. We asked two experts to segment and measure the area of the ventricle in single frames of 12 sample videos. They were instructed to do the measurement twice for each frame manually with a short break between each try. The results were 12 frames each measured 4 times. The standard deviation for each frame measurement was calculated and the average of standard deviations of the measurements in these 12 frames was about 150 pixels with 50 pixels standard deviation. This is approximately 8% of an average size ventricular area in our setting's scale. This shows the inconsistency in the manual segmentation. This could be especially significant with mutant embryos whose EF is usually very small. In most cases of the TTNtv videos, the difference of the area in the ED and ES frames is between 100 to 300 pixels (considering the resolution of the videos used in this work). However, due to the nature of neural networks, ZACAF is consistent which means that measurement of a frame multiple times will always result in only one consistent measurement.

It is noteworthy to mention since the ground truth is created using the same frames for segmentation of the ventricle, the frame rate is less important in comparing manual and automatic segmentation. The ES and ED frames are the most important frames when it comes to quantification of parameters like HR, EF, and FS. While recording the videos, the shutter of the camera takes a sequence of images with a certain fps. The higher the fps of the video, the higher chance for exact ES and ED stages being recorded. This fact cannot be proved using the metrics because the prediction is only being compared with the existing manually segmented ground truth and if the low fps causes the loss of ED or ES frames, there is no way to show it with the metrics. However, **Figure 7** shows the importance of the higher frame rate in exporting useful information, such as heartrate, and strove volume with a much better accuracy. It suggests the optimal number of fps to configure settings. Higher frame rates will result in bigger video file sizes, while low frame rate reduces the probability of identifying ED and ES frames. As seen in **Figure 7a,** a higher frame rate is particularly important with mutants since the EF is small.

From the segmentation point of view, there are two major differences between the mutant and wildtype fish. The ventricle and the heart in general have abnormal shapes in a number of mutant types. In our case here, EF is much lower in the TTNtv model as the shape as well as the contractility are significantly affected. Thus, the ventricle area difference in ES and ED frames in TTNtv mutants is very low. **Figure 8** provides examples to compare wild and TTNtv zebrafish. In some cases, the ventricle is barely beating in a way that the area difference in ED and ES frames is lower than the segmentation error. In other word, the area of the ventricle barely changes to the point that occasionally the nominator of the formula of EF is lower than the estimation error. That is the main source for the inaccuracies with the TTNtv mutants and further improvements of preprocessing or optimization of the framework will not affect the result with the mutant significantly. The videos used in this work have low resolution (data rate of 2500 kbps), in order to demonstrate the capability of our framework. Although this is beneficial for researchers to reduce required storage capacity, higher resolution would help resolve this issue,

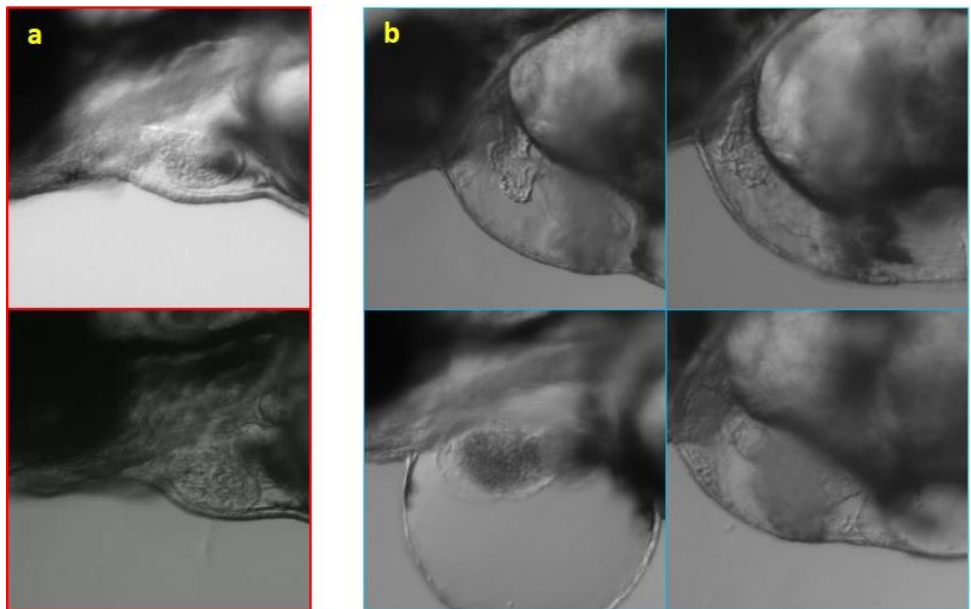

**Figure 8. Comparison of the shape and size of wildtype (a) and TTNtv mutant zebrafish (b)**. Besides the abnormal shape of the heart with the swollen ventricular wall, the smaller size of the ventricle is also found with TTNtv mutants. Further, the swollen chest can be also noticed.

thus improving the robustness and accuracy for TTNtv mutant and wildtype fish in general.

The novelty of our supervised learning-based ZACAF lies in the automatic feature and the robustness in working with black and white videos at different configurations (*i.e.* frame rate). The first novelty was demonstrated as creating mask and training can be carried out only once then ZACAF can be run easily by a non-expert person. The second novelty possesses several broader impacts. First, the versatility to work with regular bright field microscopic videos would make ZACAF widely accepted by the research community. Second, the capability to work with monochrome low fps videos would help save storage space when thousands or more videos are used. Resolution and frame rate of the videos have a direct connection with the size of them; thus, it is useful to find out the minimum required video quality. In our work, videos with resolution of 640×404 pixels, a frame rate of 5 fps, and a data rate of 2500 kbps would have a file size of about 5 megabytes for a 20-second video. It is evident that improving the resolution will increase the accuracy and robustness of the framework. This also helps reduce the processing time as well as increase the accuracy and robustness; and thus, enabling big scale projects involving multiple research groups.

For future work, we plan to not only improve the deep learning model in our ZACAF, but also include additional information in the output. It is straightforward to improve the accuracy of the deep learning model by adding more labeled data to its training dataset. The GUI can be improved in a way so the framework can process multiple videos simultaneously. We also plan to improve our framework for 3-D segmentation of the ventricle. Using the same state of the art with z-stack images of the heart sequentially recorded from the beating heart as the input allows for a 3-D segmentation of the ventricle over time. This is important because it would further avoid estimation and improve the overall accuracy significantly.

## 5. Conclusions

In this work, a framework, namely ZACAF has been developed to automatically segments the beating ventricle of zebrafish embryos from microscopic videos. The employed U-net deep learning algorithm was evaluated with wildtype and cardiomyopathy mutant fish (TTNtv) using three metrics and favorable accuracy was achieved. Our framework would help enable and accelerate numerous biological studies in cardiology and developmental biology using the zebrafish model. Moreover, as the work is being improved, it could be utilized with other animal models and even humans, and with different imaging techniques such as ultrasound or MRI imaging. Ultimately, this automated system could be translated for use not only in processing and machine learning-based analysis of various physiological parameters to support studies and disease diagnosis but also in manufacturing and automation.

**Acknowledgements**

The authors would like to acknowledge the financial support from the NSF CAREER Award #1917105 (H.C.), the NSF #1936519 (J.L. and H.C), the NIH SBIR grant #R44OD024874 (M.P.H.L and H.C.), the NIH HL107304 and HL081753 (X.X.). Also, we would like to acknowledge the GAANN fellowship support from Dept. of Education under award #P200A180052 to A.N. Lastly, we like to acknowledge the China Scholarship Council (CSC201906370239 to H.B.).